\documentclass[preprint,authoryear,12pt]{elsarticle}

\begin{document}

\begin{frontmatter}

\title{Evolution Patterns: Designing and Reusing\\ Architectural Evolution Knowledge\\ to Introduce Architectural Styles}

\author{Dalila~Tamzalit\corref{dali}, Tom Mens\corref{tom}}
\cortext[dali]{Universit\'e de Nantes, France}
\cortext[tom]{Software Engineering Lab, Universit\'e de Mons, Belgium}

\begin{abstract}
Software architectures are critical in the successful development and evolution of software-intensive systems.
While formal and automated support for architectural descriptions has been widely addressed, their evolution is equally crucial, but significantly less well-understood and supported.
In order to face a recurring evolution need, we introduce the concept of \emph{evolution pattern}. It formalises an architectural evolution through both a set of concepts and a reusable evolution process. We propose it through the recurring need of introducing an architectural style on existing software architectures. We formally describe and analyse the feasibility of architectural evolution patterns, and provide a practical validation by implementing them in \textsf{COSA\-Builder}, an Eclipse plug-in for the \textsf{COSA} architectural description language.
\end{abstract}


\end{frontmatter}

\section{Introduction}\label{sec:Introduction}

As acknowledged by \cite{Perry&al1992, ShawGarlan1996, Bass&al997, Taylor&al2009},  \emph{software architectures} have become accepted as one of the main artefacts of software development. They form an integral part of the specification of a wide variety of complex software-intensive systems. Including architectural design in the early stages of the software system life-cycle can be decisive for the software's success.
They provide a powerful abstraction mechanism that is critical to support the successful development and evolution of the software systems they describe  

The growing importance of software architecture descriptions and the maturity of the research field led to the \cite{iso2008} that defines a software architecture as \emph{``the fundamental organization of a system embodied in its components, their relationships [...] and the \textbf{principles guiding its design and evolution}.'' } 
Research on such principles for guiding architectural evolution has received relatively little attention, despite the promise of controlling cost and other change-related challenges (cf. \cite{legoaer&al2008, GarlanEtAl2009, MageeEtAl2010}).

Our main objective is to offer a disciplined way to reuse evolution knowledge within software architectures. We focus on a specific and recurring evolution need: introducing an architectural style by restructuring an existing software architecture. 
We propose to formalise and automate \emph{evolution patterns} as a means to guide and support evolution of architectural descriptions.
The overall approach follows three successive stages to specify a reusable evolution process:
(1) \emph{reify architectural concepts} that may evolve; (2) specify a minimal set of \emph{recurring evolution operations } on these concepts; and (3) specify \emph{the evolution process} through a specific workflow of evolution operations applied to identified architectural concepts. 

This article is structured as follows. Section~\ref{sec:SoftArchEvol} introduces and explains the necessary architectural concepts.  Section~\ref{sec:CaseStudySpec} presents a case study using the \textsf{COSA} architectural description language. Section~\ref{sec:GraphTransSpec} formally presents and analyses the approach using graph transformations. Section~\ref{sec:validation} explains how we implemented and validated these ideas in \textsf{COSABuilder}, an Eclipse plug-in for \textsf{COSA}. Section~\ref{sec:RelatedWork} discusses related work,  Section~\ref{sec:future} highlights some avenues of future research and Section~\ref{sec:Conclusion} concludes.

\section{Architectural concepts} \label{sec:SoftArchEvol}

The use of architectural descriptions has become well-established. They specify the various concerns of the system at a high level of abstraction. Such descriptions are made possible thanks to primary notions and concepts of viewpoints and views, architectural description languages and architectural styles.

\begin{figure}[!htb]
\centering
\includegraphics[width=0.8\textwidth]{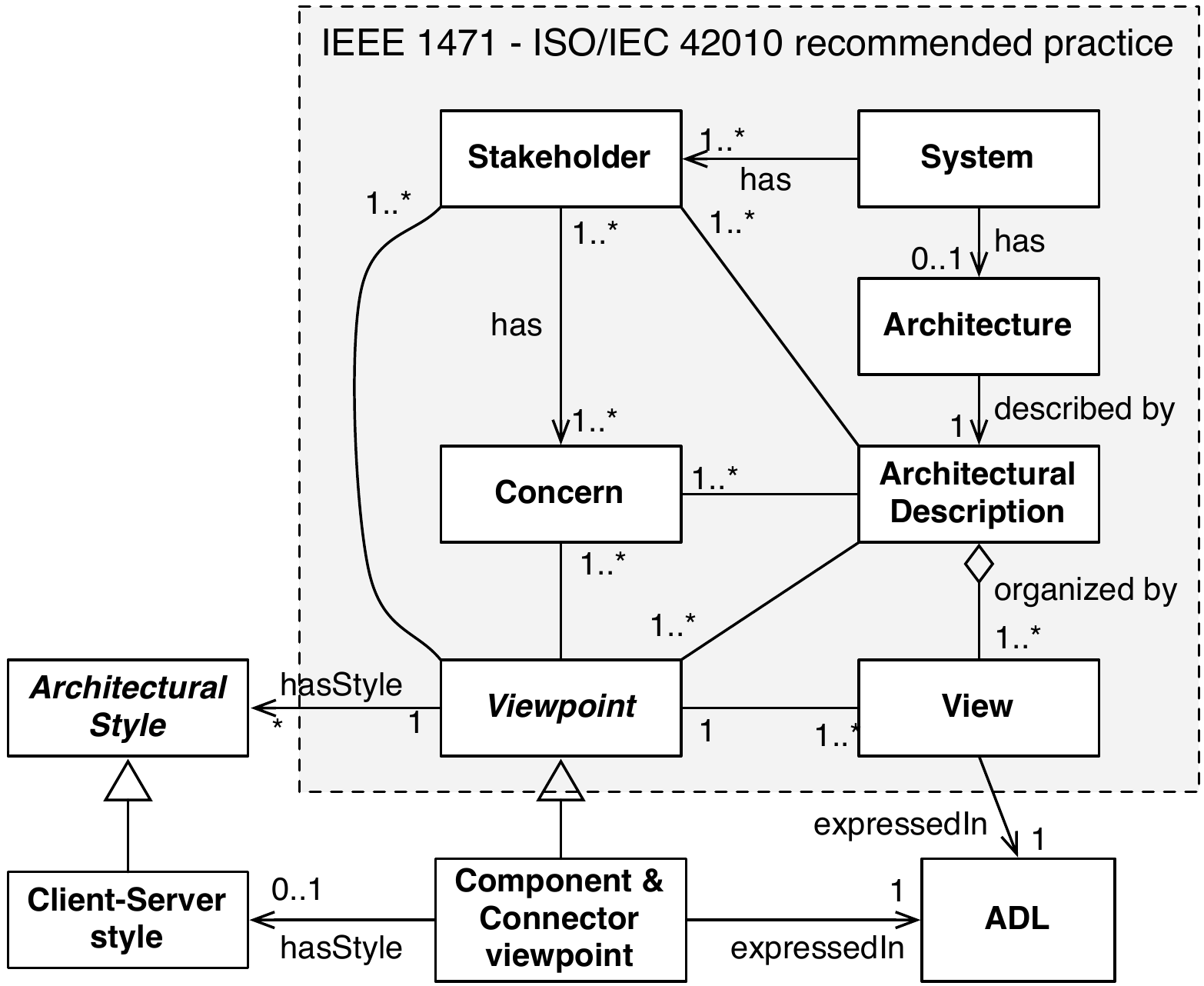}  
\caption{Fragment of the ISO/IEC standard~42010 conceptual framework enriched with the architectural style concept and a particular viewpoint.}
\label{fig:conceptualModel}
\end{figure}

The architectural description of a software-intensive system is commonly organized in several representations, like the different types of architectural blueprints in building construction. The objective is to reduce complexity and to facilitate system understanding  \cite{Clements&al2002, Kruchten1995}. We base our work on the \cite{iso2008},  partly illustrated in Figure~\ref{fig:conceptualModel} by the shaded dotted rectangle: a description of the \emph{Architecture} of a \emph{System} is composed of \emph{Views} expressed along different \emph{Viewpoints} addressing \emph{Concerns} that are important to a particular set of \emph{Stakeholders}.

Viewpoints are generic and can concern several architectures, while views are architecture-specific. Viewpoints thus allow to organise an architectural description  where each view conforms to a particular viewpoint.
According to \cite{iso2008}, one of the main viewpoints is the \emph{structural} viewpoint. Following \cite{Vestal93}, it structures the organization of an architectural description in terms of coarse-grained \emph{components} with \emph{ports} and their interactions through \emph{connectors} with \emph{roles}, ignoring technical and implementation details.
As suggested in Figure~\ref{fig:conceptualModel}, the current article focuses on structural viewpoint. We will use from now on the term  \emph{architecture} to refer to \emph{architectural description within the structural viewpoint}. 

\emph{Architectural Description Languages (ADLs)} have been proposed as a formal means to describe software architectures  following the structural viewpoint. 
Many ADLs have been proposed over the years: \emph{ACME} by \cite{Garlan&al1997}, 
\emph{Aesop} by \cite{Garlan&al1994}, 
\emph{C2SADEL} by \cite{Medvidovic&al1999}, 
\emph{Darwin} by \cite{Magee&al1995}, 
\emph{MetaH} by \cite{vestal&al1993}, 
\emph{Rapide} by \cite{Luckham&al1995}, 
\emph{SADL} by \cite{moriconi&al1997}, 
\emph{Unicon} by \cite{shaw&al1995}, 
\emph{Wright} by \cite{Allen&al1996} and 
\emph{AADL} by \cite{Lewis2002}.
Each of them has its own notation and features, often with their own supporting methods and tools. They generally propose the same concepts of \emph{component}, \emph{port}, \emph{connector} and \emph{role}. 

As advised by \cite{iso2008} and \cite{Medvidovic&Taylor2000}, an ADL should also provide support to \emph{evolve} architectural descriptions. 
However, many ADLs, generally domain-specific ones, do not support such architectural evolution. 
Those that do, typically rely on mechanisms offered by the underlying programming language, such as subtyping and inheritance or refinement. The architectural concepts that are subject to evolution are typically \emph{components} and \emph{connectors} for those ADLs that support them as first-class entities.
\cite{Shaw&al1996} introduced the concept of \emph{architectural style} as a disciplined mechanism to guide the design and use of architectures. \cite{Clements&al2002, Garlan&al1993, Garlan&al1994} refer to an \emph{architectural style} as a family of architectures in terms of a \emph{pattern of structural organization} through a coordinated set of architectural constraints. These architectural constraints define: a unified vocabulary of component and connector types; constraints on relations between these types; and a semantic interpretation for each instantiated element.
\cite{Gomaa&al1998} rely on architectural styles to facilitate the \emph{construction} of architectures, while \cite{Garlan&al1994, Shaw&al1996} use styles to constrain and evolve architectures. 

Among the best known structural architectural styles are the \textit{Pipe-and-Filter} style of \cite{Garlan&2002} and the \emph{Client-Server} style of \cite{Clements&al2002}. Any of these architectural styles defines specific types of components, ports, connectors and roles in addition to a set of architectural constraints. 
In this article we focus on the \emph{Client-Server}  architectural style as a case study. Figure~\ref{fig:conceptualModel} shows how to fit it into the ISO/IEC 
conceptual framework. 

Evolution mechanisms proposed by ADLs are generally tied to supporting tools, and thus hardly reusable when a similar architectural evolution situation occurs. 
An explicit specification of \emph{evolution pattern} would enable future reuse, thereby reducing the costs and risks of architectural evolution in the long run. A typical example of such a reoccurring pattern is the restructuring of a monolithic architecture of a legacy system into a distributed client-server architecture. 

\emph{Evolution patterns} can be specified in terms of more elementary predefined architectural \emph{evolution operations}. While an elementary evolution operation can  lead a given architecture to an inconsistent state, an evolution pattern allows to evolve an architecture from a consistent state to another consistent one. As an evolution pattern should be reusable, its well-formedness needs to be validated before being proposed for reuse in similar evolution situations on different architectures. Upon application of the evolution pattern, automatic analyses can be performed to determine the conformity of the evolved architecture, and to report or resolve any conformance problems that may arise.

\section{Case Study}\label{sec:CaseStudySpec}

\subsection{The \textsf{COSA} ADL}\label{sec:COSA-ADL}

In this article, we will illustrate our approach through the specification, analysis and execution of an evolution pattern to introduce the Client-Server architectural style on a monolithic e-shop architecture.
To do so, we will rely on the \textsf{COSA} ADL developed by \cite{Maillard2007} and its associated tool \textsf{COSABuilder}.
We use \textsf{COSA} for four main reasons:
\emph{(i)} it is generic and extensible, defined through a metamodel; 
\emph{(ii)} it manipulates architectural elements (configuration, component, connector, port, role, \ldots) as first-class entities; 
\emph{(iii)} it can be easily extended with new first-class concepts, which is very useful if we want to add evolution operations and evolution patterns;
\emph{(iv)} the metamodel and source code of \textsf{COSABuilder} are available to us.

The notion of computation (represented by \emph{components}) is separated from the notion of interaction and communication (represented by \emph{connectors}). A component has a set of ports (provided or required). The topological structure of the architecture is represented within a \emph{configuration}, a graph of interconnected components. Each composite component has its own configuration that handles its internal architectural elements. Connectors can either be user-defined or built-in. In the latter case, we distinguish between \emph{attachments} (to connect a port to a role) and \emph{bindings} (to interconnect two ports or two roles, generally in case of delegation).

\subsection{The e-shop architecture}\label{sec:EShopArchitecturalView}

\begin{figure*}[!htb]
 \begin{center}  
        \includegraphics[width=13cm]{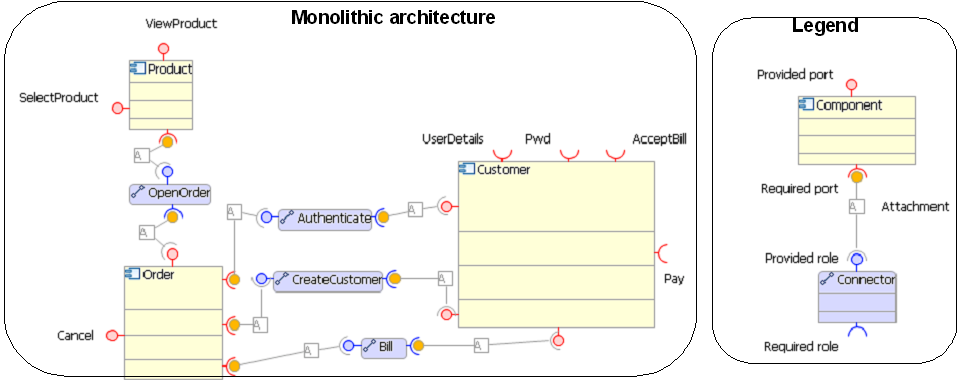}   
         \caption{The EShop monolithic architecture expressed using the \textsf{COSA} ADL.}         
         \label{fig:EShopArchitectureSimpleSpecNoDep}
\end{center}	
\end{figure*}

Figure~\ref{fig:EShopArchitectureSimpleSpecNoDep} presents the monolithic e-shop architecture in \textsf{COSA}. 
It contains three main components: \texttt{Product}, \texttt{Customer} and \texttt{Order} with their own ports in addition to a set of connectors.
We only explain one component, \texttt{Customer}, as the two other  components follow the same spirit. \texttt{Customer} has four required ports (\texttt{UserDetails}, \texttt{Pwd}, \texttt{AcceptBill} and \texttt{Pay}) and three provided ports (\texttt{Authenticate}, \texttt{CreateCustomer} and \texttt{Bill}). For aims of clarity, the names of these provided ports are hidden since the connectors they are attached to have the same names.
 
The e-shop architecture also specifies how its components interact together through four connectors: \texttt{OpenOrder}, \texttt{Authenticate}, \texttt{CreateCustomer} and \texttt{Bill}. Each of these connectors has two roles, to connect a provided port of one component to a required port of another component. Some component's ports, such as \texttt{UserDetails} are connection-free and represent the interaction points of e-shop with its environment. 

\subsection{Extending \textsf{COSA} with structural dependencies}\label{sec:InternalDependencies} 

Since we focus on the structural viewpoint in this article, we restrict ourselves to expressing \emph{structural dependencies}. 
Ideally, architectural restructuring, expressed by evolution patterns, should preserve dependencies: if some provided port of a component (transitively) depends on a required port of a (possibly different) component, this should remain the case after the restructuring. 

In order to address properly architectural restructuring, we need to consider all structural dependencies between ports of components. These dependencies can be of two kinds. \emph{External dependencies} between different components are expressed using connectors, attachments and bindings. \emph{Internal dependencies} between ports of  the same component are generally not explicitly  materialized. We enriched the \textsf{COSA} metamodel and syntax with a new built-in connector type, named \emph{uses (U)}, to represent these internal dependencies. The \emph{Product} component in Figure~\ref{fig:SplitComp}\textbf{(a)} has two such internal dependencies: port \texttt{SelectProduct} \emph{uses} port \texttt{ViewProduct} (a customer first needs to view products in order to select one), and port \texttt{OpenOrder} \emph{uses} port \texttt{SelectProduct} (a customer can only order selected products).
Although not explicitly shown in Figure~\ref{fig:EShopArchitectureSimpleSpecNoDep} for aims of readability, the e-shop architecture is enriched with these internal structural dependencies.

\subsection{Introducing the Client-Server architectural style}\label{sec:CSStyle}
\label{sec:StructureDep}

Architectural restructurings, such as the migration to a Client-Server architecture typically need to preserve internal dependencies.
Generally used for network-based applications, the Client-Server style proposes two additional types of component, \emph{Server} and \emph{Client}, that are connected together. A software architecture conforming to the Client-Server style is only allowed to have instances of the element types specified by the style. In addition, it must respect all constraints imposed by the style. 
A server component offers a set of services to its client(s). A client component, desiring for a service to be performed, may send a request to the server via a connector. The server either rejects or performs the request and sends a response back to the client.
In this article, we will adopt the following variant:
there must be exactly one server within a given architecture and at least one client, and each client must be connected to at least one server. In addition, any other type of component must be contained (possibly indirectly) in either a client or the server.

\begin{figure*}[!htb]
\centering
\includegraphics[width=\textwidth]{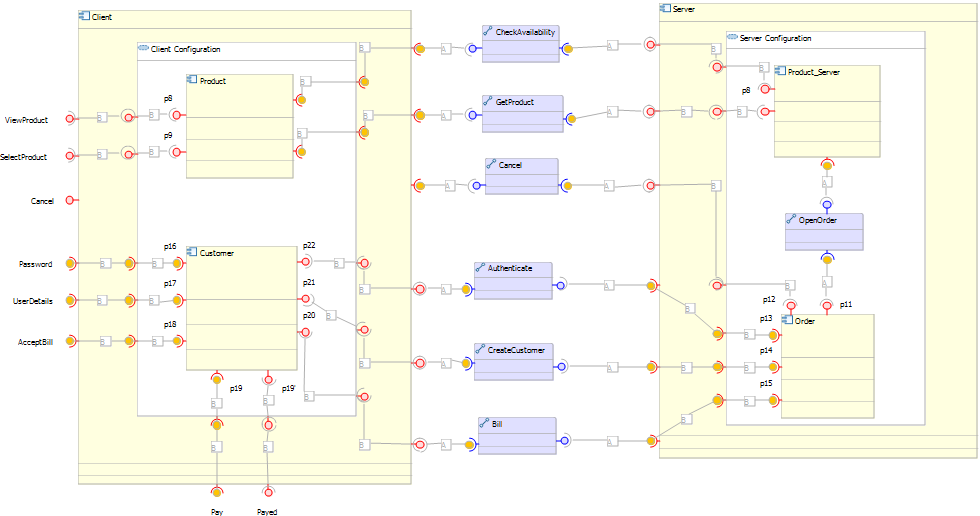}
\caption{Resulting client-server e-shop architecture in \textsf{COSA} after automatic changes and specific changes triggered by the architect.} 
\label{fig:ClientServerArchitectureVer2}
\end{figure*}

\begin{figure*}[!htb]
\centering
\includegraphics[width=13.5cm]{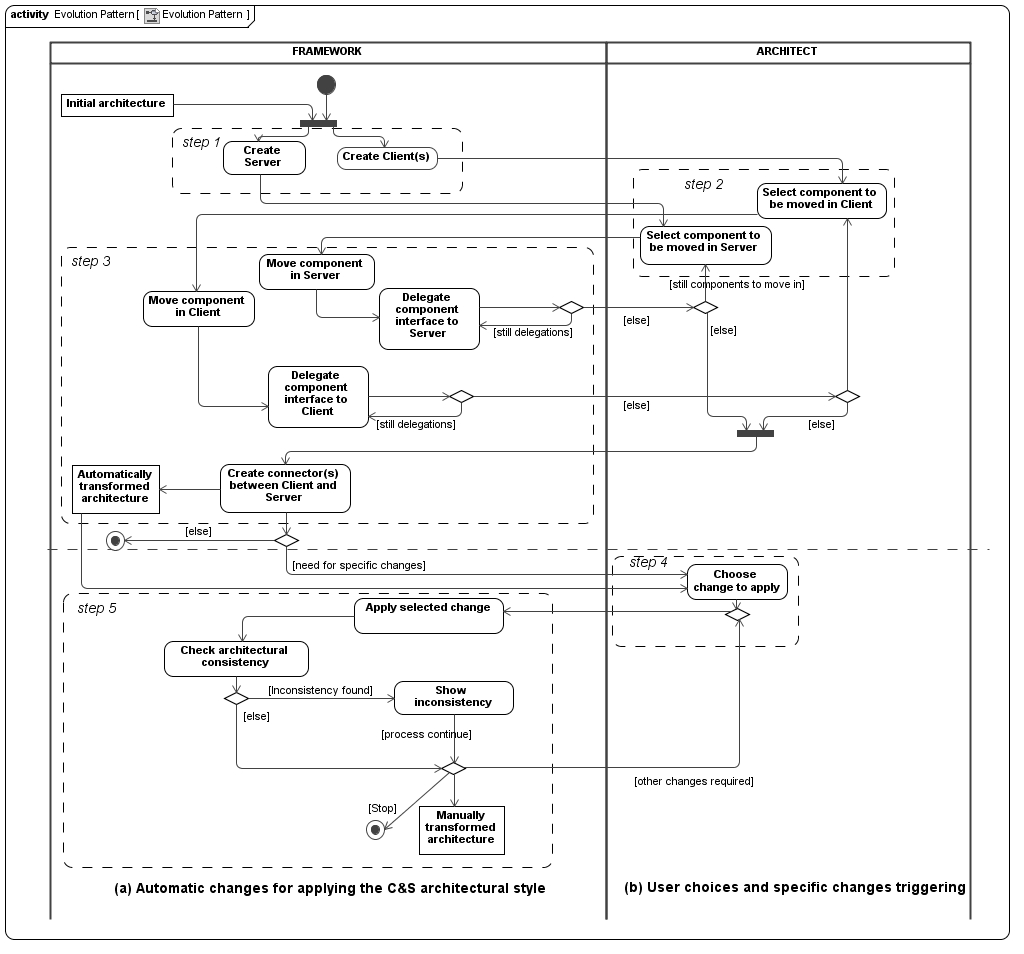}
\caption{Evolution pattern representing the introduction of the Client-Server architectural style. (In the activities \emph{Delegate component interface to Server} (respectively \emph{Client}) we used the more generic term interface instead of port.)}
\label{fig:ActivityDiag}
\end{figure*}

Our goal is now to specify an \emph{evolution pattern} that evolves the e-shop architecture of Figure~\ref{fig:EShopArchitectureSimpleSpecNoDep}
into the architecture of Figure~\ref{fig:ClientServerArchitectureVer2} that conforms to the Client-Server style.
The evolution pattern represents an  architectural restructuring that preserves all external ports and structural dependencies of the original architecture.
Figure~\ref{fig:ActivityDiag} shows the evolution patterns as a UML activity diagram, with two swimlanes representing (a) the automated changes carried out by the \emph{framework}; and (b) the changes manually triggered by the \emph{architect} and executed by the framework. The activity diagram is composed of five basic steps, indicated by dashed rounded rectangles:

\begin{enumerate}

\item The \emph{framework} creates one server and $n$ clients ($n\geq1$) with the names of all new components specified by the architect. For our e-shop architecture, $n=1$, so one \texttt{Server} component and one \texttt{Client} component are created.

\item As client(s) and the server are the only top-level components allowed, the architect selects existing components to be moved into a client or into the server.  To obtain Figure~\ref{fig:ClientServerArchitectureVer2}, the architect chooses to move the \texttt{Order} component in \texttt{Server}, and the \texttt{Product} and \texttt{Customer} components in \texttt{Client}.

\item The \emph{framework} moves the selected components in the server and client(s) and transforms automatically all connectors and dependencies consistently to remain conform to the \textsf{COSA} ADL and the Client-Server architectural style. 

\item The architect manually triggers further desired changes. In the example, she wishes to have a \texttt{Product} component in both \texttt{Server} and \texttt{Client} with different ports. It is thus necessary to \emph{split} the \texttt{Product} component in two components, \texttt{Product} (that allows the user to view and select products) and \texttt{Product\_Server} (that encapsulates product information in the server). The architect also wishes to provide the \texttt{Cancel} service from the \texttt{Server}.

\item The \emph{framework} executes the requested changes and checks whether the architecture still conforms to the Client-Server style and the \textsf{COSA} ADL. In the example, splitting \texttt{Product} and moving the resulting component  \texttt{Product\_Server} in \texttt{Server} requires to adapt existing dependencies and connectors, yielding the final architecture in Figure~\ref{fig:ClientServerArchitectureVer2}.
\end{enumerate}

In section~\ref{sec:GraphTransSpec} we show how this evolution pattern can be formally specified and analysed, and section~\ref{sec:validation} shows how it is implemented in the \textsf{COSABuilder} tool.

\subsection{Extending COSA with Evolution Operations}\label{sec:EvolOp}

The \emph{evolution pattern} of Figure~\ref{fig:ActivityDiag} is essentially defined as an application of elementary architectural evolution operations in a particular order. We have identified many such operations in \textsf{COSA}, such as: \emph{Move Port} from a component to another, \emph{Split Component} into two or more components, \emph{Merge Components} into one component, \emph{Move in Component} as a subcomponent of another and \emph{Move out Component} from its containing component.

To restructure the e-shop architecture into a Client-Server one, we applied several of these evolution operations.
During step 3 of Figure~\ref{fig:ActivityDiag}, \emph{Move in Component}  was used to move components into \texttt{Client} or \texttt{Server}, followed by transformation \emph{Delegate Component Port} to create necessary bindings and connectors using more primitive transformations like \emph{Move Port} coupled with the creation of corresponding ports and attachments on \texttt{Client} and \texttt{Server}. 
During step 5, the \emph{Split component} operation was applied to achieve specific changes requested by the user: splitting the \texttt{Product} component (belonging to \texttt{Client}) in two components \texttt{Product} and \texttt{Product\_Server}. The latter one is subsequently \emph{moved out} of the \texttt{Client} to be \emph{moved in} to the \texttt{Server}.

\begin{figure*}[!htb]
\centering
\includegraphics[width=\textwidth]{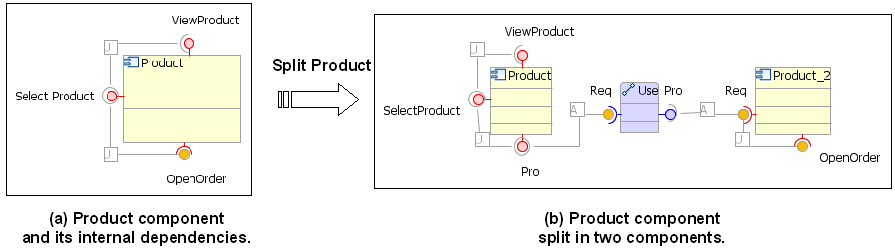}
\caption{\emph{Split Component} restructuring. (\textbf{U}-rectangles represent \emph{uses} dependencies.)}
\label{fig:SplitComp}
\end{figure*}
 
This interactive evolution operation is guided and restricted by structural constraints, as illustrated in Figure~\ref{fig:SplitComp}.
\emph{Split Component} starts by selecting a component (here: \emph{Product}), creating one or more new components (here: \emph{Product\_2}\footnote{This name is automatically generated but can be changed by the architect.}), and moving some manually selected ports of \emph{Product} into \emph{Product\_2}. 
While moving ports, in order to preserve structural dependencies we need to take into account three different situations:

\begin{itemize}
\item If the port to be moved does not depend on any other port, it can be moved to the target component without any additional changes. 

\item If two ports with a \emph{uses}-dependency between them are both moved to the same target component, the internal dependency is moved along. 

\item If the port to be moved has a \emph{uses}-dependency to or from another port, this dependency needs to be preserved by the evolution pattern. This is for example the case in Figure~\ref{fig:SplitComp} for \emph{Open Order} port that uses \emph{Select Product}.
To preserve the dependency, after moving \emph{OpenOrder} to \emph{Product\_2}, a provided port \emph{Pro} on \emph{Product} and a required port \emph{Req} on \emph{Product\_2} are created, and a connector is added to connect these two ports.
In addition, \emph{uses}-dependencies are respectively created between the new port \emph{Pro} and \emph{SelectProduct} and the new port \emph{Req} and \emph{Open Order}. As a result, the initial \emph{uses} dependency from \emph{Open Order} to \emph{Select Product} is automatically replaced by a longer dependency path between both ports, via an intermediate connector and two newly created ports and two newly created \emph{uses}-dependencies. 
\end{itemize}

\section{Formalising and Analysing Architectural Evolution} \label{sec:GraphTransSpec}  

A prerequisite for providing automatic support for architectural evolution is the ability to formally specify the ADL, architectural styles and evolution patterns. We use graph transformation theory for this purpose. The analogy between architectural evolution and graph transformation is quite natural: an architecture description can be expressed as a graph containing a set of interconnected components.  
 Graph transformations allow us to formally analyse and reason about architectural evolution operations.

In this section we explain how we use AGG\footnote{\texttt{http://user.cs.tu-berlin.de/\~}\texttt{gragra/agg/}} for this puropse, a Java-based graph transformation engine conceived by \cite{Taentzer2003AGG}. We exploit AGG's built-in formal analysis mechanisms to reason about evolution patterns and their relation to the architectural styles.

\subsection{Formalising the \textsf{COSA} ADL} \label{sec:TypeGraph} 

\begin{figure*}[!htb]
 \begin{center}  
\includegraphics[width=\textwidth]{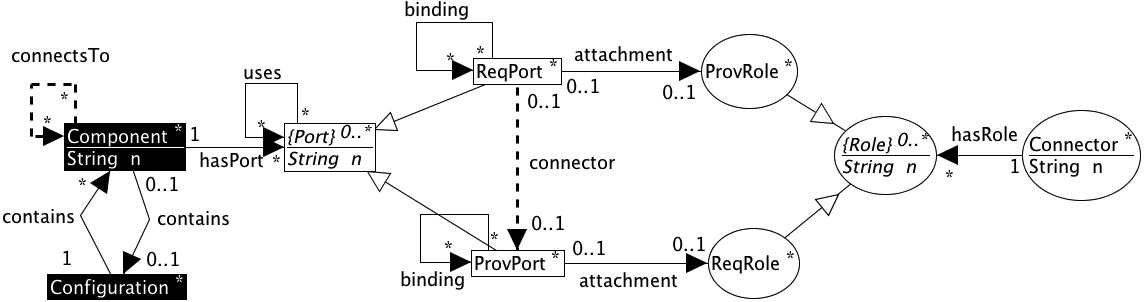}
\caption{Type graph for a part of the \textsf{COSA} metamodel.}
\label{fig:typeGraph}
\end{center}
\end{figure*}

Figure~\ref{fig:typeGraph} shows how the part of the \textsf{COSA} metamodel that is of interest to us is expressed as a \emph{type graph}.
 Any well-formed architecture can be represented as a graph that conforms to this type graph.
The type graph represents all architectural concepts we want to reason about (e.g., Component, Port, Connector, Role) as a  \emph{node type}. The associations or relations between the architectural concepts (e.g., bindings, attachments, uses dependencies and containment relationships) are represented by \emph{edge type}s.

The type graph imposes (lower and upper) multiplicity constraints on edge types. Node and edge types may contain additional attributes, and inheritance can be used between node types, as explained by \cite{deLara2007AGT}. This is the case between the abstract node type \emph{Port} (resp. \emph{Role}) and its two concrete subtypes \emph{ProvPort} (resp. \emph{ProvRole}) and \emph{ReqPort} (resp. \emph{ReqRole}) that represent provided and required ports (resp. roles). All node types contain an attribute \texttt{n} of type \emph{String} to represent the name of the corresponding node. 
A \emph{Port} (resp. \emph{Role}) should always be connected by an edge of type \emph{hasPort} (resp. \emph{hasRole}) to exactly one \emph{Component} (resp. \emph{Connector}), and a \emph{Component} (resp. \emph{Connector}) may have any number of \emph{Ports} (resp. \emph{Roles}).
The edge type \emph{contains} relates a component to one of its subcomponents (via an intermediate \emph{Configuration} node type). The \emph{binding} edge type represents the binding of a port of the component to a port (of the same type) belonging to its subcomponent.
The \emph{uses} edge type represents a structural internal dependency between ports.

\begin{figure}[!htb]
 \begin{center}  
\includegraphics[width=6.8cm]{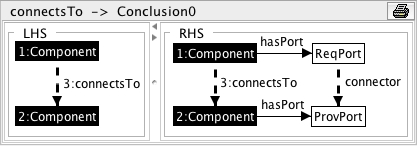}\vspace{0.5cm}
\includegraphics[width=6.5cm]{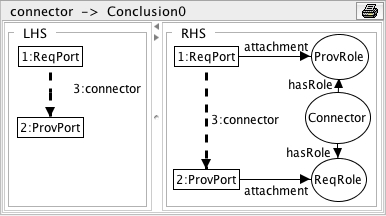}
\caption{Derived edge types \emph{connector} and \emph{connectsTo} are expressed in terms of a path of other edge types and node types (shown on the right).}
\label{fig:graphInvariant0}
\end{center}
\end{figure}

In order to simplify the formal representation of architectures, the type graph also includes so-called ``derived" edge typess (\emph{connectsTo} and  
 \emph{connector}. Figure~\ref{fig:graphInvariant0} shows how they are expressed in terms of a more complex path of other node types and edge types.

\begin{figure}[!htb]
 \begin{center}  
\includegraphics[width=6.9cm]{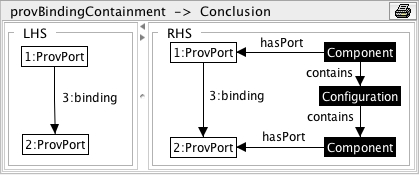}\hspace{0.5cm}
\includegraphics[width=6cm]{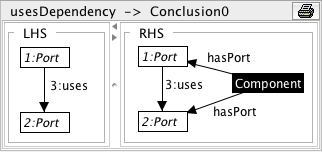}

\caption{Graph invariants representing additional well-formedness constraints on the type graph.}
\label{fig:bindingContainment}
\end{center}
\end{figure}

In addition to the type graph, graph invariants are needed to specify well-formedness constraints that cannot be expressed directly by the type graph. Some of these are: (i) a component cannot be connected to, or contained in,  itself;
(ii) a \emph{binding} is only allowed between ports of the same type belonging to a component and one of its subcomponents;
(iii) a \emph{uses} dependency is only allowed between different ports belonging to the same component;
(iv) two components cannot be at the same time connected to, and contained in, one another.
Constraint (ii) and (iii) are formally expressed as graph invariant in Figure \ref{fig:bindingContainment}. The others can be expressed in a similar way.

\begin{figure*}[!htb]
\centering
\includegraphics[width= \textwidth]{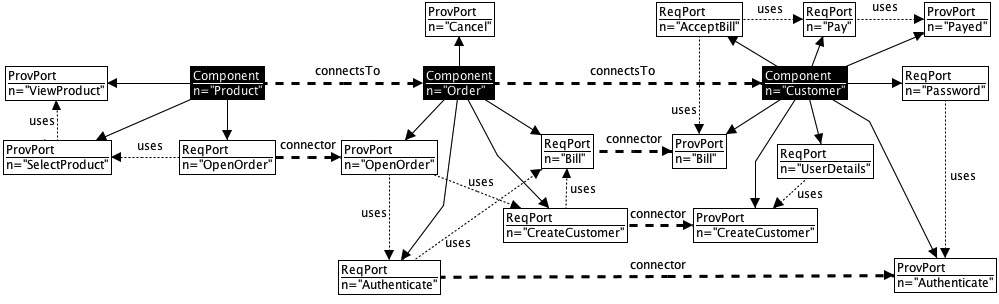}
\caption{A graph representing the abstract syntax of the initial e-shop architecture.}
\label{fig:graph}
\end{figure*}

A \textsf{COSA} architecture can be represented as a \emph{graph} conforming to the type graph together with all of its graph invariants.
Figure~\ref{fig:graph} shows the graph corresponding to the e-shop architecture of Figure~\ref{fig:EShopArchitectureSimpleSpecNoDep}. 
This time, we have explicitly shown the \emph{uses}-dependencies that were kept hidden in Figure~\ref{fig:EShopArchitectureSimpleSpecNoDep}. With AGG, we can automatically verify that this graph conforms to its type graph and that all imposed graph invariants are satisfied. 

\subsection{Formalising architectural styles} \label{sec:ArchStyles} 
To formalise an architectural style,
 we proceed in a similar way as for formalising the \textsf{COSA} ADL: we extend the type graph and add graph invariants that express the additional constraints imposed by the architectural style\footnote{Other important information that can typically be found in a style guide \cite{Clements&al2002}, such as the rationale behind the style and the ``what it is for and not for'' section, have not been formalised in this article. How these aspects of a style can be expressed formally is left as a topic for future work.}.

\begin{figure}[!htb]
\centering
\includegraphics[width=6.5cm]{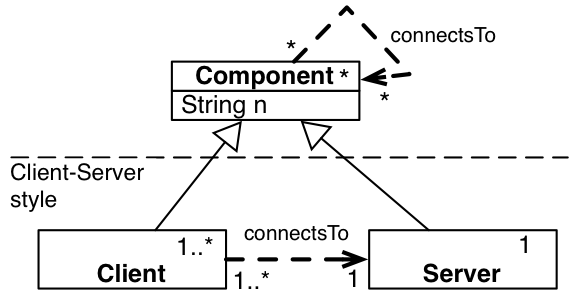}\hspace{0.5cm}
\includegraphics[width=6cm]{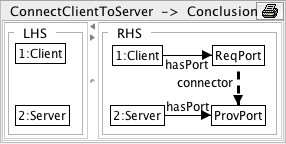}
\caption{Extension of the \textsf{COSA} type graph to accommodate the Client-Server architectural style. The \emph{connectsTo} association between \emph{Client} and \emph{Server} refines (i.e., constrains) the one on \emph{Component}.}
\label{fig:CSTG}
\end{figure}

The left part of figure~\ref{fig:CSTG} shows how two new \emph{Component} subtypes need to be added to the type graph  to represent the Client-Server architectural style : a \emph{Client} node type and a \emph{Server} node type.
The node multiplicities state that there should always be one server and at least one client. The edge multiplicities state that each client must be connected to one server.

In addition, we need to add two extra graph invariants that further constrain the type graph:
(i) a \emph{Client} component must always connected to \emph{Server} via one of its ports (see right part of figure~\ref{fig:CSTG});
(ii) any component that is not a \emph{Client} or \emph{Server} must be contained in another component
(i.e., only \emph{Client} and \emph{Server} are allowed as top-level components).
For the sake of simplicity, we have not modeled the use of a particular client-server protocol. How to specify and analyse such a protocol is a topic of future work.

A distinct advantage of formalising architectural styles is the ability to verify whether a given architecture is well-formed (i.e., it conforms to its ADL), and whether it conforms to an architectural style. This kind of verification is quite straightforward. We have defined the \textsf{COSA} ADL and the Client-Server architectural style as a type graph together with additional graph constraints, and AGG provides direct support for checking whether a graph conforms to its type graph and its associated graph constraints. This checking was successfully done on our e-shop architecture, before and after introduction of the Client-Server style.

\subsection{Formalising evolution pattern} \label{sec:GraphTransformation} 
\label{sec:RuleSequences}

In order to specify architectural evolution patterns we rely on the notion of a \emph{graph transformation}. Essentially, a graph transformation takes a graph as input and produces another graph as output. It is specified by means of a \emph{graph transformation rule} that must conform to the type graph and all graph constraints.

\begin{figure}[!htb]
\centering
\includegraphics[width=8cm]{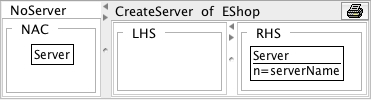}\vspace{0.2cm}
\includegraphics[width=9.5cm]{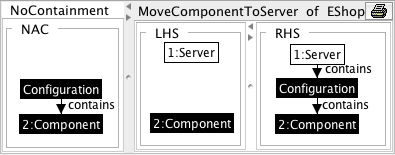}\vspace{0.2cm}
\includegraphics[width=12cm]{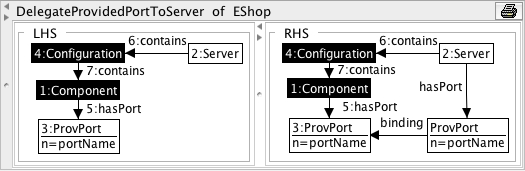}
\caption{Three graph transformation rules that are part of the evolution pattern to move a component into a server. Next to \emph{DelegateProvidedPortToServer} we defined a similar transformation rule for \emph{DelegateRequiredPortToServer}.}\label{fig:moveToServer}
\end{figure}

Figure~\ref{fig:moveToServer} shows three examples of graph transformation rules that formalise activities of Figure~\ref{fig:ActivityDiag}: \emph{Create Server}, \emph{Move Component To Server} and \emph{Delegate Provided Port to Server}. 
 In general, the specification of a graph transformation rule is composed of three different parts (displayed from left to right in the figure): a number of optional negative application conditions (NAC), a left-hand side (LHS) and a right-hand side (RHS). Applying the transformation proceeds as follows:
\begin{enumerate}
\item an occurrence (or ``match'') of the LHS needs to be found in the host graph. It is possible that multiple matches are found. In that case, the user selects the desired match or the tool chooses a match non-deterministically.
\item if successful, the NAC is used to verify that certain ``forbidden constructs'' do not appear in the match. In rule \emph{CreateServer} of figure~\ref{fig:moveToServer}, NAC \emph{NoServer} states that a server cannot be created if there is already one\footnote{In contrast, multiple clients are allowed.}. NAC \emph{NoContainment} of rule \emph{MoveComponentToServer} states that the component to be moved is not allowed to be contained in any other configuration.
\item if the NAC is satisfied, the transformation rule is applied by ``replacing'' the match corresponding to the LHS in the host graph by its RHS. Identical numbers in LHS and RHS are used to identify nodes and edges that are to be preserved by the transformation. Nodes or edges only appearing in the RHS are newly added; nodes or edges only appearing in the LHS are removed by the transformation.
\end{enumerate}

\emph{Evolution patterns} can be formalised, in part, as sequences of graph transformation rules.
The rule sequence below specifies the order in which to apply the transformation rules of Figure~\ref{fig:moveToServer} to move components into the server:\\
 {\small \texttt{CreateServer; (MoveComponentToServer$)^*$;\\
(DelegateProvPortToServer$)^*$; (DelegateReqPortToServer$)^*$}}\\
The \texttt{;} symbol specifies the order in which to apply the rules, whereas the \texttt{*} is used to specify a repetition of a rule (or even a sequence of rules), by applying it as long as a new match can be found in the graph.
Rule sequences provide a way to formalise the \emph{evolution pattern} of Figure~\ref{fig:ActivityDiag} that specifies the introduction of the Client-Server architectural style. An activity diagram represents a set of possible execution paths. Each such path can be expressed by a rule sequence. As such, the set of rule sequences represents all possible execution scenarios specified by the evolution pattern.

AGG provides different ways to analyse whether an evolution pattern is well-defined, including \emph{critical pair analysis (CPA)} and \emph{rule sequence analysis}.
CPA is used to detect parallel conflicts and sequential dependencies between pairs of transformation rules $R_1$ and $R_2$. \emph{Parallel conflicts} represent situations in which two transformation rules are not jointly applicable to the same host graph: application of rule $R_1$ prohibits subsequent application of $R_2$ or vice versa. This is used, for example, to verify whether two rules are mutually exclusive. \emph{Sequential dependencies} represent situations where rule $R_2$ is causally dependent on $R_1$: $R_2$ cannot be applied directly to the host graph, but becomes applicable once $R_1$ has been applied.

As an example, consider the evolution pattern of Figure~\ref{fig:ActivityDiag}. Starting from an initial software architecture, one can execute two parallel sequences of activities\footnote{The fork notation in Figure~\ref{fig:ActivityDiag}, denoted by a black horizontal bar with two outgoing transitions, represents the start of two parallel threads of execution.}. The first sequence starts with \emph{Create Server} followed by \emph{Move Component to Server} and \emph{Delegate (Required or Provided) Port to Server} (step 3).
 The second sequence is similar but for Client instead of Server.

\begin{figure}[!htb]
 \begin{center}  
\includegraphics[width=11cm]{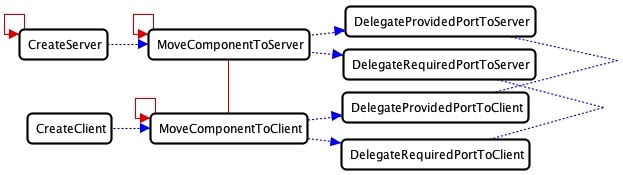}
\caption{Critical pair analysis of  graph transformation rules. Dotted arcs represent sequential dependencies, solid arcs represent parallel conflicts.}
\label{fig:CPA-results}
\end{center}
\end{figure}

To verify that the imposed order between the different activities actually makes sense, we specified each of them as a graph transformation rule, and ran the critical pair analysis to detect parallel conflicts and sequential dependencies between pairs of rules. The result of this analysis is shown in Figure~\ref{fig:CPA-results}.
It corresponds to our intention: \emph{Create Client} and \emph{Create Server} are parallel independent; they can be applied in parallel without any harm. \emph{Create Server} is in parallel conflict with itself because only one server can be introduced into the client-server architecture. \emph{Move Component to Client} and \emph{Move Component to Server} have a potential parallel conflict if one tries to move the same component in the client and in the server.
\emph{Move Component to Server (resp. Client)} is in conflict with itself because one cannot move the same component twice. We also find all expected sequential dependencies: \emph{Delegate (Provided or Required) Port to Server} causally depends on \emph{Move Component to Server} that causally depends on \emph{Create Server}.
This automated formal analysis makes us more confident that the evolution pattern specified in Figure~\ref{fig:ActivityDiag} for applying the Client-Server architectural style is well-defined. In fact, we used the formal analysis as a kind of debugging mechanism: it enabled us to fine-tune the graph transformation rules and the well-formedness constraints imposed by the architectural style.

\begin{figure*}[!htbp]
 \begin{center}  
\includegraphics[width=\textwidth]{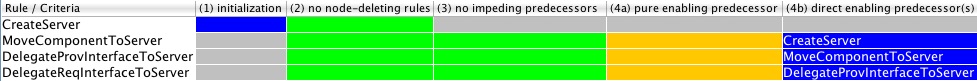}
\caption{Automated applicability analysis of the rule sequence that specifies how to move components into the server.} 
\label{fig:ruleseq-anal}
\end{center}
\end{figure*}

As explained by \cite{Lambers&al2008} and \cite{Jurack&al2008}, it is possible to analyse rule sequences and this analysis is supported by AGG. We can specify any given path in an activity diagram as a rule sequence, verify whether this sequence is applicable and, if not, what are the potential applicability problems. For the example rule sequence defined earlier, Figure~\ref{fig:ruleseq-anal} shows the result of analysing applicability (for details, see \cite{Lambers&al2008}). 
It reveals that the analysed rule sequence does not have any particular inconsistency (it would have been displayed in red otherwise). As a result, the application of the rules in the specified order will not provoke conflicts. 
This confirms the results concerning sequential dependencies observed in Figure~\ref{fig:CPA-results}. The rule sequence for moving components into a server thus is well-defined. Analysing the rule sequence to move components into a client yields a similar result.

\section{Automating Architectural Evolution}\label{sec:validation}

\begin{figure*}[!htb]%
\includegraphics[width=\textwidth]{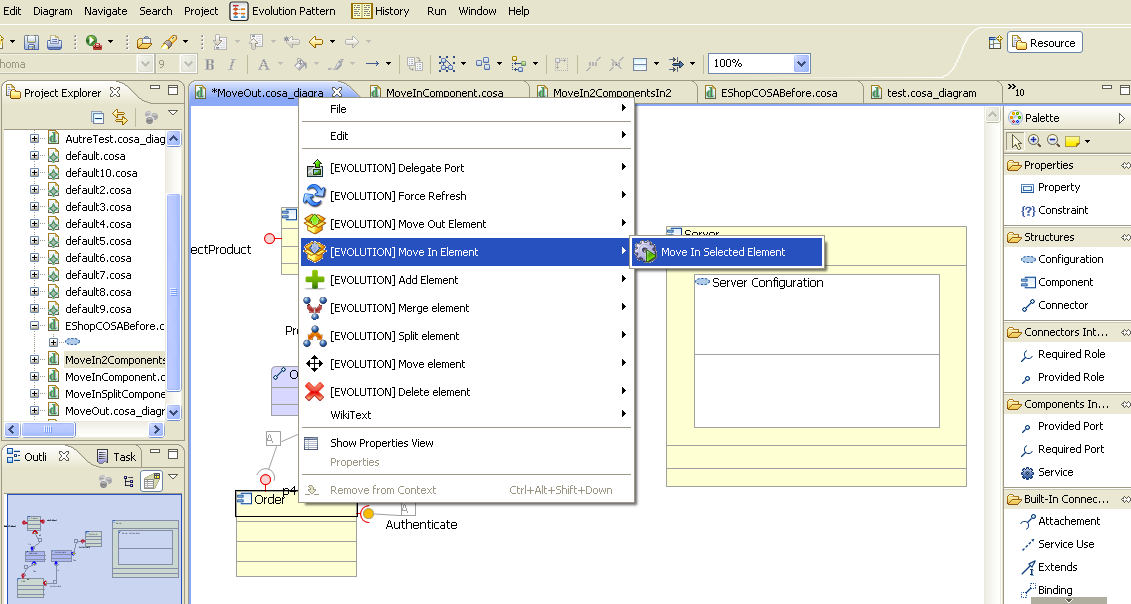}
\caption{Extension of \textsf{COSABuilder} tool with support for architectural evolution.}%
\label{fig:COSAEvolFramework}%
\end{figure*}

\begin{figure*}[!htb]%
\includegraphics[width=0.95\textwidth]{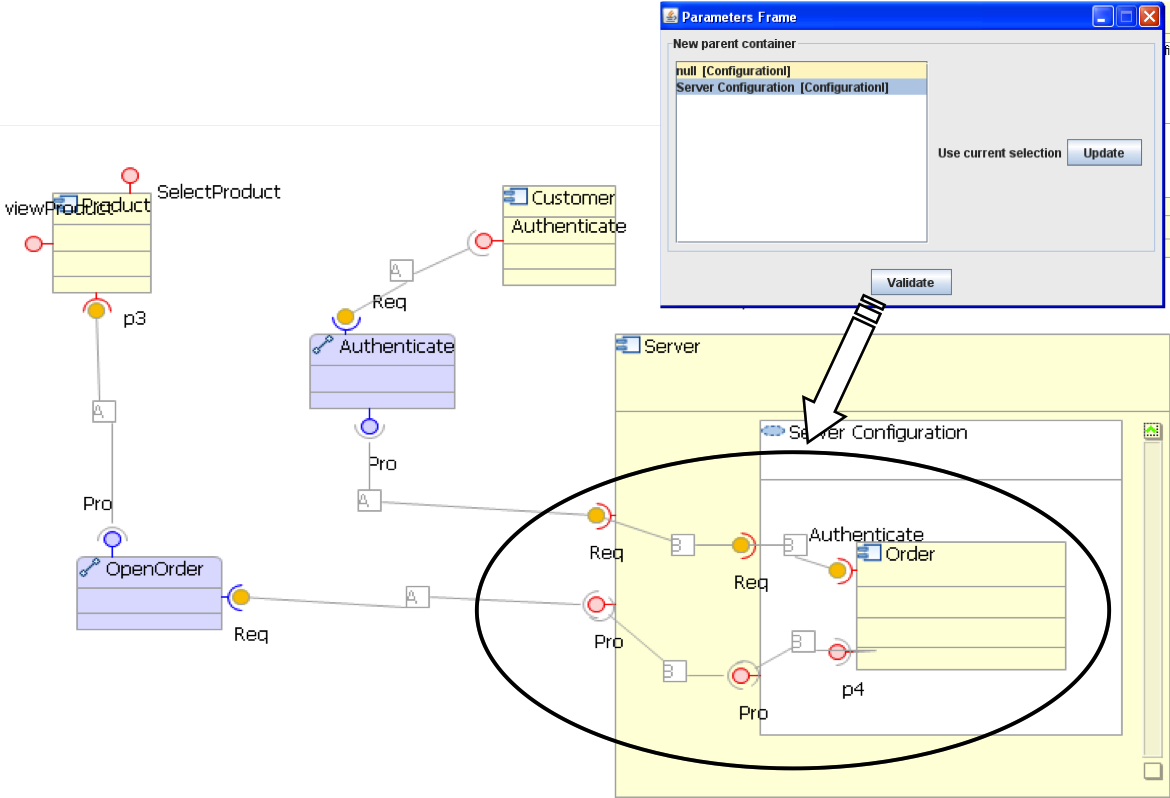}%
\caption{Result of applying the the \emph{Move In} evolution operation in \textsf{COSAEvol}.}%
\label{fig:COSAEvolMoveIn}%
\end{figure*}

After this formal validation of our ideas, we present now the practical validation through the extension of \emph{COSABuilder}, an Eclipse plug-in that implements the \textsf{COSA} ADL to specify architectural descriptions. Its extension with support for evolution patterns and architectural styles is shown in Figure~\ref{fig:COSAEvolFramework}. 
The right frame displays the initial palette of \textsf{COSABuilder} enriched with the concept of \emph{Service Use} to represent uses dependencies between ports. The central frame displays the architecture description. Evolution operations can be applied to it by selecting the appropriate architectural element and choosing the desired evolution operation from a context-sensitive menu. Figure~\ref{fig:COSAEvolFramework} shows the \emph{Move In} operation on the selected \emph{Order} component. Figure~\ref{fig:COSAEvolMoveIn} shows a new menu that appears (to select the target component into which \emph{Order} needs to be moved), as well as the architecture resulting from applying this evolution operation.

The implemented evolution layer enables the definition of: \emph{(i)} elementary architectural \emph{evolution operations}, \emph{(ii)} reified \emph{evolution patterns} in terms of those elementary evolution operations and \emph{(iii)} \emph{architectural styles} that can be checked. 
Evolution operations and evolution patterns are defined as first-class entities, by extending the \textsf{COSA} meta-model with two new metaclasses: \emph{EvolutionOperation} and \emph{EvolutionPattern}.

Each evolution operation is implemented as a Java class using the \emph{Singleton} design pattern. It has a unique method \emph{execute(parameters)}. The first parameter is the \emph{context}: the architectural element to be modified. Other parameters are specific to the evolution operation under consideration.
We implemented the following elementary evolution operations, in such a way that they preserve internal \emph{uses} dependencies:
\begin{itemize}
	\item \emph{Create} or \emph{Delete} an architectural element (Component or Connector) or interface (Port or Role). 
	\item \emph{Move} an architectural element to a new parent.
	\item \emph{Split} an architectural element into two or more.
	\item \emph{Merge} several architectural elements into one. 
	\item \emph{Move Out} an element from its containing element, or \emph{Move In} an element inside another one (shown in Figures~\ref{fig:COSAEvolFramework} and~\ref{fig:COSAEvolMoveIn}).
	\item \emph{Delegate}: create a binding from a given port to another element in the architecture, generally its parent. 
\end{itemize}

\begin{figure}[!htb]
 \begin{center}  
\includegraphics[width=11cm]{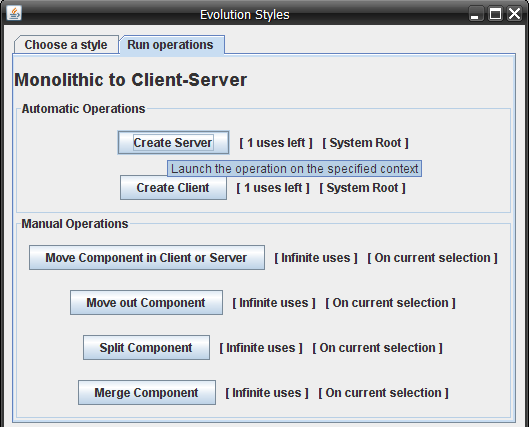}
\caption{Running implementation of the Client-Server evolution pattern.}
\label{fig:COSABuilder-style}
\end{center}
\end{figure}

Evolution patterns can be implemented in terms of existing evolution operations. Figure~\ref{fig:COSABuilder-style}
illustrates the execution of the evolution pattern of Figure~\ref{fig:ActivityDiag} to introduce the Client-Server style.
 
\begin{figure}[!htb]
 \begin{center}  
\includegraphics[width=11cm]{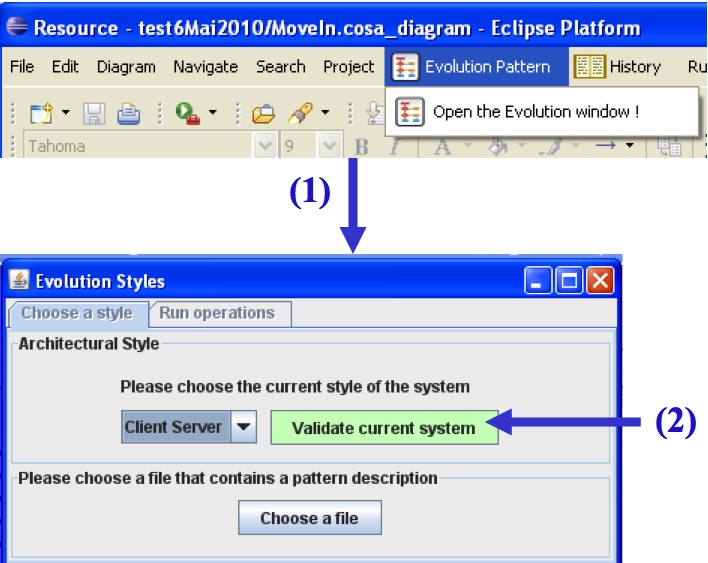}
\caption{Checking conformance to the Client-Server architectural style.}
\label{fig:COSACheckStyle}
\end{center}
\end{figure}

\begin{figure}[!htb]
 \begin{center}  
\includegraphics[width=6cm]{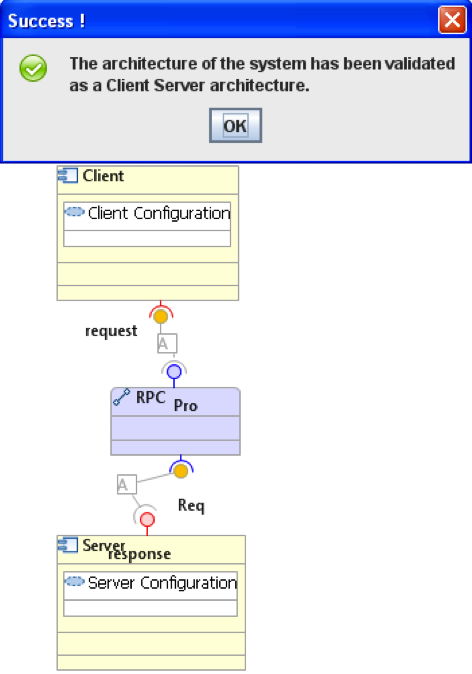}
\includegraphics[width=6cm]{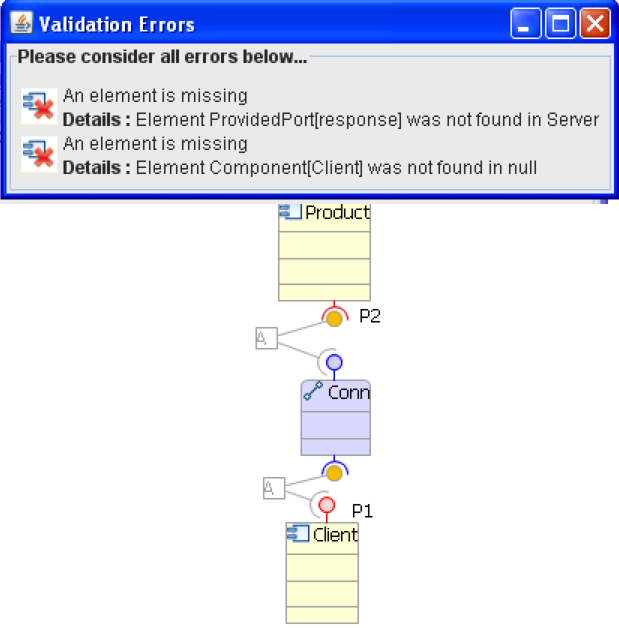}
\caption{Result of checking conformance to the Client-Server architectural style on two different architectures.}
\label{fig:COSACheckStyleResults}
\end{center}
\end{figure}

Figure~\ref{fig:COSACheckStyle} illustrates how to check conformance of an architecture to the Client-Server architectural style. This is achieved in three steps: (1) open the Evolution window; (2) select the architectural style to be checked on the architecture; and (3) display the result of applying the architectural style, together with a message window indicating success or failure of the conformance checking. Figure~\ref{fig:COSACheckStyleResults} shows an example of success on the left, and an example of failure on the right.

\section{Related Work} \label{sec:RelatedWork}
A variety of related work focuses on the individual topics that form part of our architectural evolution framework.
The novelty of our research is that we combine several topics together by introducing the use of evolution patterns, formally specifying and analysing them using graph transformation to introduce and check architectural styles, and implementing them in an architectural modeling tool for the \textsf{COSA} ADL.

\cite{lemetayer1998} used graph transformation theory to describe software architecture and architectural style as graph grammars. He proposed to use a coordinator in terms of rules that must be statically checked, as opposed to our use of evolution patterns that introduce architectural styles.
\cite{WermelingerFiadeiro2002} used graph transformation theory to present an algebraic foundation for software architecture reconfiguration. 
\cite{Grunske2005} formalised architectural refactorings as hypergraph transformation rules that can be applied automatically.

The idea of considering architectural styles as typical architectural ``evolution patterns'' was first introduced by \cite{tamzalit&al2006a} who proposed the SEM model that introduces the concept of \emph{evolution style}. \cite{Garlan2008} defined an evolution style as a set of evolution paths among different kinds of systems. 

\cite{McVeighEtAl2006} provide first-class operations that express and capture architectural changes. The main difference is the type of evolution operations that are provided and the ADL for which they are supported. An additional difference is that we use these evolution operations as elementary building blocks to create more complex evolution patterns to introduce architectural styles.

\cite{Noppen&al2010} go beyond the evolution pattern by proposing a framework to tailor evolution processes according to some desired architectural traits by looking for them in a given architectural knowledge base. 

\section{Future perspectives}\label{sec:future}

The current article only scratched the surface of how one should provide more disciplined support for architectural evolution. This section provides a roadmap of future work that is still required in this very important emerging research domain.

Our formalisation of the structural viewpoint and the Client-Server architectural style, only focused on the architectural elements, relations and structural constraints between them.  As explained by \cite{Medvidovic&Taylor2000}, many ADLs enable the formal analysis and verification of  important non-functional properties such as consistency, completeness, correctness, performance, reliability, security, availability and dependability. We need to integrate support for verifying such properties, and to preserve these properties during architectural evolution. We also need support for documenting the rationale behind an architecture and its imposed architectural styles.

Because \textsf{COSA} offers 4 layers of modeling (similar to the OMG MDA architecture), the tool support can accommodate other ADLs as well by considering \textsf{COSA} as a pivot ADL.
\cite{Smeda2005} proposed strategies to transform an architectural description specified in \textsf{COSA} to a description in another ADL such as UML 2.
This would allow us to benefit from a wide range of tools that have been implemented for the UML language.

Our case study presented only one viewpoint and architectural style. Applying it to other viewpoints and styles opens up the possibility to evolve an architectural description that involves multiple viewpoints, and to deal with architectural styles that span multiple viewpoints. 
Other interesting scenarios are the replacement of an architectural style by another one on an existing architecture.

Another future research topic is the study of \emph{co-evolution} between ADLs, architectural styles, their conforming architectures, and their implementation. Changes to any of these entities may require changes to the others. We thus need to analyse this change impact and manage co-evolution while limiting the number of constraint violations. This becomes very challenging when multiple architectural styles co-exist that may interfere when applied to the same software architecture.

\section{Conclusion} \label{sec:Conclusion}

In this article, we introduced architectural evolution patterns as a disciplined mechanism to introduce architectural styles to an architectural description. We provided a case study using the structural viewpoint and the Client-Server architectural style expressed in the \textsf{COSA} ADL. Using graph transformation, we formally analysed evolution patterns by relying on the notions of critical pair analysis and rule sequence analysis. This allowed us to ensure that the evolution patterns we specified are well-formed and preserve the consistency constraints imposed by the ADL and the architectural style. We implemented and practically validated our ideas by extending \textsf{COSABuilder}, the tool that accompanies the \textsf{COSA} ADL, with explicit and first class support for defining and applying evolution patterns and architectural styles.\\

\bibliographystyle{elsarticle-num-names}

\end{document}